\documentclass[aps,
pre,
superscriptaddress,
nobibnotes,
reprint,
twocolumn,
longbibliography,
]{revtex4-2}

\usepackage{graphicx} 
\usepackage[utf8]{inputenc}
\usepackage[T1]{fontenc}
\usepackage{mathptmx}
\usepackage{amsmath,amsthm,amssymb,bm}
\usepackage{xcolor}

\begin{document}

\title{Spatial correlations of entangled polymer dynamics}

\author{Jihong Ma}
\altaffiliation{Current address: Department of Mechanical Engineering, University of Vermont, Burlington, VT 05405, USA}
\affiliation{Center for Nanophase Materials Sciences, Oak Ridge National Laboratory, Oak Ridge, Tennessee 37831, USA}

\author{Jan-Michael Y. Carrillo}
\affiliation{Center for Nanophase Materials Sciences, Oak Ridge National Laboratory, Oak Ridge, Tennessee 37831, USA}

\author{Changwoo Do}
\affiliation{Neutron Scattering Division, Oak Ridge National Laboratory, Oak Ridge, Tennessee 37831, USA}

\author{Wei-Ren~Chen}
\affiliation{Neutron Scattering Division, Oak Ridge National Laboratory, Oak Ridge, Tennessee 37831, USA}

\author{P\'{e}ter~Falus}
\affiliation{Institut Laue-Langevin, B.P. 156, F-38042 Grenoble CEDEX 9, France}


\author{Zhiqiang Shen}
\affiliation{Center for Nanophase Materials Sciences, Oak Ridge National Laboratory, Oak Ridge, Tennessee 37831, USA}

\author{Kunlun Hong}
\affiliation{Center for Nanophase Materials Sciences, Oak Ridge National Laboratory, Oak Ridge, Tennessee 37831, USA}

\author{Bobby G. Sumpter}
\affiliation{Center for Nanophase Materials Sciences, Oak Ridge National Laboratory, Oak Ridge, Tennessee 37831, USA}

\author{Yangyang Wang}
\email{wangy@ornl.gov}
\affiliation{Center for Nanophase Materials Sciences, Oak Ridge National Laboratory, Oak Ridge, Tennessee 37831, USA}

\begin{abstract}
The spatial correlations of entangled polymer dynamics are examined by molecular dynamics simulations and neutron spin-echo spectroscopy. Due to the soft nature of topological constraints, the initial spatial decays of intermediate scattering functions of entangled chains are, to the first approximation, surprisingly similar to those of an unentangled system in the functional forms. However, entanglements reveal themselves as a long tail in the reciprocal-space correlations, implying a weak but persistent dynamic localization in real space. Comparison with a number of existing theoretical models of entangled polymers suggests that they cannot fully describe the spatial correlations revealed by simulations and experiments. In particular, the strict one-dimensional diffusion idea of the original tube model is shown to be flawed. The dynamic spatial correlation analysis demonstrated in this work provides a useful tool for interrogating the dynamics of entangled polymers. Lastly, the failure of the investigated models to even qualitatively predict the spatial correlations of collective single-chain density fluctuations points to a possible critical role of incompressibility in polymer melt dynamics.
\end{abstract}

\date{\today}

\maketitle

\section{INTRODUCTION}
The topological constraints arising from chain connectivity and excluded volume forces have a profound influence on the dynamical properties of macromolecular liquids~\cite{porter1966entanglement,graessley1974entanglement,lodge1992dynamics}. Despite the extensive research efforts of the past several decades, notably around the framework of the tube model~\cite{de1971reptation,DE1,DEbook}, the precise physical nature of polymer entanglement is still not fully understood. In this work, we demonstrate a fruitful approach to the dynamics of entangled polymers by calling attention to the spatial correlations of their density fluctuations. In a typical neutron spin-echo spectroscopy (NSE) experiment or computer simulation, intermediate scattering functions are studied at discrete wavenumbers. Here, we show that an ``orthogonal'' approach --- probing spatial correlations of dynamics at different correlation times --- can provide valuable insights into the molecular motions of entangled polymers. 

To illustrate this idea, we start by discussing our molecular dynamics simulations (MD) on entangled and untangled polymer melts, performed with the GPU-accelerated LAMMPS package~\cite{LAMMPS,plimpton1995fast,brown2011implementing}. We consider a semi-flexible coarse-grained bead-spring model~\cite{grest2016communication}, where the non-bonded interactions are described by an attractive Lennard-Jones potential with cutoff $r_{\mathrm{c}}=2.5\sigma$ and the beads along the polymer chains are connected by FENE bonds. The chain stiffness is controlled by a bending potential $U_{\mathrm{bend}}=k_{\theta}(1+\cos\theta)$, where $\theta$ is the angle between two successive bonds and $k_{\theta}=1.5$ in the current study. Three different systems with chain length $N=25$, $400$, and $2000$ are simulated at a reduced density $\rho=0.89$ and temperature $T=1$. Since the entanglement length $N_{\mathrm{e}}\approx 28$~\cite{grest2016communication}, the $N=25$ melt is unentangled, whereas the $N=400$ and $2000$ systems are entangled. To describe the microscopic dynamics, we focus on two time correlation functions: the incoherent intermediate scattering function $S_{\mathrm{inc}}(Q,t)$ and the normalized single-chain dynamic structure factor $S(Q,t)/S(Q)$. Additional information of the simulations and data analysis is described in Sections \ref{sec:IIA} and \ref{sec:IIB}.

\begin{figure}
    \centering
    \includegraphics[width=\columnwidth]{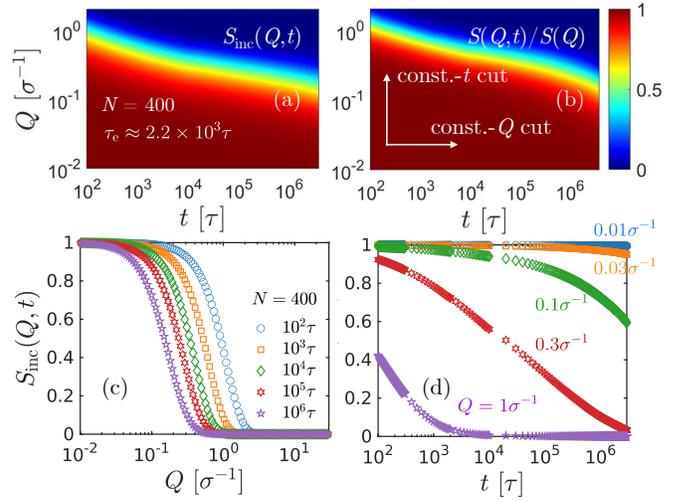}
    \caption{(a) 2D color map of the incoherent dynamic structure factor $S_{\mathrm{inc}}(Q,t)$ of the $N=400$ semi-flexible system simulated by MD. Both $Q$ and $t$ are presented in reduced Lennard-Jones units. (b) Corresponding normalized single-chain dynamic structure factor $S(Q,t)/S(Q)$. (c) Spatial dependence of $S_{\mathrm{inc}}(Q,t)$ at different correlation times ($t=10^2$, $10^3$, $10^4$, $10^5$, and $10^6\tau$). (d) Temporal correlation of $S_{\mathrm{inc}}(Q,t)$ at different wavenumbers ($Q=1$, $0.3$, $0.1$, $0.03$, and $0.01\sigma^{-1}$).}
    \label{fig:MD_illustration}
\end{figure}

Representative 2D spatiotemporal maps~\cite{shen2021spatiotemporal} of $S_{\mathrm{inc}}(Q,t)$ and $S(Q,t)/S(Q)$ are shown in Figs.~\ref{fig:MD_illustration}a and \ref{fig:MD_illustration}b for the $N=400$ melt. The traditional method in this field~\cite{kremer1990dynamics,ewen1997neutron,putz2000entanglement} places emphasize on analyzing time correlations of scattering functions at discrete wavenumbers. An alternative approach, however, is to ask how the \textit{spatial correlations of dynamics} are affected by entanglements on different time scales by taking cuts along lines of constant correlation time on the 2D landscape (Fig.~\ref{fig:MD_illustration}). To contrast this proposed approach with the traditional method, an example is given in Figs.~\ref{fig:MD_illustration}c and \ref{fig:MD_illustration}d for the incoherent scattering function $S_{\mathrm{inc}}(Q,t)$. The functional forms of $S_{\mathrm{inc}}(Q,t)$ appear to be similar, when compared at different correlation times (Fig.~\ref{fig:MD_illustration}c). On the other hand, $S_{\mathrm{inc}}(Q,t)$ does not retain its shape, when examined at different wavenumbers. This distinction between the two approaches can be intuitively appreciated by treating Figs.~\ref{fig:MD_illustration}a and \ref{fig:MD_illustration}b as topographic maps and imaging ourselves as travelers exploring the corresponding terrains. Walking along the $Q$ direction (cuts of constant correlation time), our experience is always similar --- the slopes resemble one another (Fig.~\ref{fig:MD_illustration}c). By contrast, the trail is highly dependent on the choice of $Q$, if we travel along the $t$ direction. In other words, we find simplicity in the spatial correlation analysis. 

Moreover, there are deeper theoretical considerations for taking this alternative point of view of polymer dynamics. According to the reptation idea~\cite{de1971reptation,DE1}, entanglement interactions effectively confine a tagged polymer in a tube-like region. We may reason, therefore, that such a topological confinement must leave fingerprints on the spatial correlations of chain motions. Indeed, on time scales shorter than the disengagement time $\tau_{\mathrm{d}}$, the classical tube model~\cite{DE1,fatkullin1995theory} predicts that the incoherent scattering function can be described by the following equation:
\begin{equation}
    S_{\mathrm{inc}}(Q,t)\approx \exp[Q^4y(t)]\mathrm{erfc}[Q^2\sqrt{y(t)}],\label{eq:tubeSinc}
\end{equation}
where $\mathrm{erfc}(x)$ is the complementary error function, and $y(t)=a^2\langle s^2(t)\rangle/72$, with $a$ being the tube diameter and $\langle s^2(t)\rangle$ the mean-square curvilinear segment displacement. While the temporal decay of $S_{\mathrm{inc}}(Q,t)$ at a given $Q$, according to Eq.~(\ref{eq:tubeSinc}), takes on a complicated functional form, its spatial dependence at a constant correlation time is much simpler:
\begin{equation}
    S_{\mathrm{inc}}(Q,t)\approx \exp(x^2)\mathrm{erfc}(x),\label{eq:functionForm}
\end{equation}
where $x = Q^2C_t^2$, with $C_t$ a constant determined by $t$. Eq.~(\ref{eq:functionForm}) produces a \textit{heavy-tailed} distribution, in stark contrast to the Gaussian behavior of Rouse chains (see Sec~\ref{sec:TheoreticalResults} for details). The physical picture behind this prediction is that the real-space localization causes the dynamic correlations to spread out in the reciprocal space. Conversely, following this line of thinking, one should be able to characterize the entanglement effect by examining the spatial correlations of intermediate scattering functions.

In this context, it is helpful to mention the pioneering work of Granick and coworkers~\cite{wang2010confining}, who determined the tube confining potential for entangled F-actin solutions using single-molecule fluorescence imaging. Their approach was based on analysis of probability distribution of transverse displacement relative to the \textit{tube axis}. Our proposal of spatial correlation analysis for the classical scattering functions, which are defined in the \textit{laboratory frame}, is philosophically connected to the idea of Granick \textit{et al.}, but more general --- it applies to both self and collective density fluctuations.    

With these insights, this work sets out to systematically explore the benefits of spatial correlation analysis for understanding entangled dynamics, using both molecular dynamics simulations and neutron spin-echo spectroscopy experiments. The results are analyzed and contrasted with several representative theoretical models of polymer dynamics, including the Rouse model~\cite{rouse1953theory}, standard tube model~\cite{de1971reptation,DE1,deGennes1981coherent,DEbook,fatkullin1995theory}, slip-spring model~\cite{likhtman2005single}, Ronca model~\cite{ronca1983frequency}, and des Cloizeaux model~\cite{desCloizeaux1993dynamic}. The remainder of this paper is organized as follows. The details of the computational and experimental methods employed in this work are described in Section~\ref{sec:II}. Section~\ref{sec:MDresults} presents spatial correlation analyses of the intermediate scattering functions from MD simulations. Section~\ref{sec:NSEresults} reports results from NSE experiments on entangled polyethylene melts. In Section~\ref{sec:TheoreticalResults}, theoretical predictions of spatial correlations of dynamics are examined and compared with the experimental and computational results. Section~\ref{sec:relations} discusses the relation of the present MD and NSE studies with those in the literature. The final section summarizes the main conclusions of this investigation and outlines a few key implications.

\section{Computational and experimental methods}\label{sec:II}
\subsection{Dynamic spatial correlation analysis}\label{sec:IIA}
The idea of dynamic spatial correlation analysis is a simple one: direct examination of the wavenumber dependence of intermediate scattering functions (ISFs)~\footnote{In this paper, the terms intermediate scattering function and dynamic structure factor are used synonymously. Following the convention of the polymer physics community, $S_{\mathrm{inc}}(Q,t)$ is mostly referred to as incoherent (intermediate) scattering function, whereas $S(Q,t)$ is referred to as single-chain dynamic structure factor.} at fixed correlation times. In the current coarse-grained molecular dynamics simulations, the \textit{single-chain} dynamic structure factor (intermediate scattering function) $S(\mathbf{Q},t)$ can be evaluated as 
\begin{equation}
    S(\mathbf{Q},t)=\frac{1}{N}\sum_{i,j}^{N}\langle  \exp[i\mathbf{Q}\cdot(\mathbf{R}_j(t)-\mathbf{R}_i(0))]\rangle,\label{eq:SqtDef}
\end{equation}
where $N$ is the total number of beads in a chain, $\mathbf{Q}$ is the wavevector, and $\mathbf{R}_j(t)$ is the position of the \textit{j}th bead at time $t$. Similarly, the incoherent intermediate scattering function $S_{\mathrm{inc}}(\mathbf{Q},t)$ can be calculated as
\begin{equation}
    S_{\mathrm{inc}}(\mathbf{Q},t)=\frac{1}{N}\sum_{j}^{N}\langle  \exp[i\mathbf{Q}\cdot(\mathbf{R}_j(t)-\mathbf{R}_j(0))]\rangle.\label{eq:SincDef}
\end{equation}
Experimentally, the normalized single-chain dynamic structure factor $S(Q,t)/S(Q)$ can be obtained from neutron spin-echo measurements of properly labeled polymer melts. While it is straightforward to investigate the spatial correlations of intermediate scattering functions at discrete correlation times, experimental and computational studies of polymer dynamics have so far placed most attention to the time correlation of ISFs at particular wavenumbers. It is worth noting that spatial correlation analysis of the self van Hove function $G_{\mathrm{s}}(\mathbf{r},t)$~\cite{van1954correlations} (often referred to as probability density function or propagator), which is the spatial Fourier transform of $S_{\mathrm{inc}}(\mathbf{Q},t)$, is a standard practice in the literature of Brownian motions~\cite{chandrasekhar1943stochastic,metzler2000random}. From this perspective, the current study is a natural extension of this existing approach. The scope of this investigation is nevertheless confined to spatial correlations of polymer dynamics in the Fourier space. The results of real-space correlation analysis will be reported in a future publication.

\subsection{Molecular dynamics simulations}\label{sec:IIB}
\subsubsection{Semi-flexible chain model}
Most of the simulation results reported in this work are based on a coarse-grained semi-flexible bead-spring model. The non-bonded interaction between beads is described by the LJ potential,
\begin{equation}
U_\textrm{LJ}(r)=\begin{cases} 4\epsilon[(\frac{\sigma}{r})^{12}-(\frac{\sigma}{r})^{6}]-4\epsilon[(\frac{\sigma}{r_{\mathrm{c}}})^{12}-(\frac{\sigma}{r_\textrm{c}})^{6}]& r<r_\textrm{c} \\
0 & r\ge r_{\mathrm{c}} \\
\end{cases},\label{eq:LJ}
\end{equation}
with a cutoff distance $r_{\mathrm{c}}=2.5\sigma$~\cite{grest2016communication}. A finitely extensible nonlinear elastic (FENE) potential coupled with purely repulsive Weeks-Chandler-Andersen (WCA) potential~\cite{weeks1971role} is used to connect two neighboring  beads along a polymer chain:
\begin{equation}\label{eq:FENE_WCA}
U_\textrm{FENE} (r)=-\frac{1}{2}kR^2_0\ln[1-(r/R_0)^2]+U_{\mathrm{WCA}}(r),
\end{equation}
where $R_0=1.5\sigma$ and $k=30\epsilon/\sigma^2$.  Additionally, a bond-bending potential is considered:
\begin{equation}\label{eq:bending}
 U_\textrm{bend}(\theta)=k_\theta(1+\cos\theta),
\end{equation}
where $\theta$ is the angle between two subsequent bonds and $k_\theta=1.5\epsilon$. Each chain has $N$ beads of mass $m$. Equilibrium molecular dynamics simulations were performed at a reduced density $\rho=0.89$ and temperature $T=1$, and three different chain lengths were investigated: $N=25$, $400$, and $2000$.

\subsubsection{Fully flexible chain model}
To complement the molecular dynamics simulations of the semi-flexible chain model, we further consider a fully flexible bead-spring model~\cite{kremer1990dynamics,Cao2015simulating,Xu2018,lam2018scaling}. In this model, the pair interactions between any two beads are described by the WCA potential, \textit{i.e.}, the repulsive part of the Lennard-Jones potential [Eq.~(\ref{eq:LJ}) with $r_{\mathrm{c}}=2^{1/6}\sigma$]. The bond connectivity along the chain is maintained by the FENE potential, with $R_0=1.5\sigma$ and $k=30\epsilon/\sigma^2$. Equilibrium molecular dynamics simulations of melts of $N=40$ and $N=500$ were carried out at a reduced density $\rho=0.85$ and temperature $T=1$. Both the semi-flexible and fully flexible chain simulations were performed at the Oak Ridge Leadership Computing Facility with the GPU-accelerated LAMMPS package~\cite{LAMMPS,plimpton1995fast,brown2011implementing}.

\subsection{Slip-spring simulations}
In Section~\ref{sec:TheoreticalResults} of this paper, we compare the results of our molecular dynamics simulations and neutron-spin echo experiments with a number of theoretical models, including the slip-spring model of Likhtman~\cite{likhtman2005single}. This model is numerically solved by the Brownian dynamics (BD) simulation technique. In our simulations, the average entanglement spacing $N_{\mathrm{e}}$ is $4$ and the spring strength $N_{\mathrm{s}}$ is set to $1/2$. The ratio of the friction coefficient of the slip-link $\zeta_{\mathrm{s}}$ to that of the chain segment $\zeta$, $\zeta_{\mathrm{s}}/\zeta$, is 0.1. In Likhtman's work, the basic time unit $\tau_0$ is $\tau_0=\zeta b^2/\big(3\pi^2 k_{\mathrm{B}} T\big)$, with $b$ being the segment size. On the other hand, our choice of time unit is $\tau_0 = \zeta b^2/(2k_{\mathrm{B}}T)$, which is more common for Brownian dynamics simulations. The reduced simulation time is defined as $t^{*}=t/\tau_0$. To simplify the simulation, the non-crossing condition is not implemented for the slip-links. As pointed out by Likhtman~\cite{likhtman2005single}, such a choice makes only a very small difference in the results. To benchmark with the simulations of Likhtman, we choose the same set of chain lengths: $N=8$, $16$, $32$, $64$, and $128$ and verify that the shear relaxation modulus $G(t)$ in Ref.~\cite{likhtman2005single} can be correctly reproduced. The intermediate scattering functions of the $N=64$ and $128$ systems are analyzed and presented in this paper.

\subsection{Neutron spin-echo spectroscopy}

To complement the molecular dynamics investigation, neutron spin-echo spectroscopy experiments were performed on an isotopically labeled polyethylene (PE) melt with 10~v\% \textit{h}-PE ($M_{\mathrm{w}}=79$ kg/mol, $M_{\mathrm{w}}/M_{\mathrm{n}}=1.05$) and 90~v\% \textit{d}-PE ($M_{\mathrm{w}}=89$ kg/mol, $M_{\mathrm{w}}/M_{\mathrm{n}}=1.05$) at 470 and 509~K on the IN15 beamline of the Institut Laue-Langevin. A wavelength of $\lambda=10\ \textup{\AA}$ with $\Delta\lambda/\lambda=15\%$ was used to achieve a measurement range of $0.021\le Q\le 0.21\ \textup{\AA}\mathrm{^{-1}}$ and $0.019\le t\le 194\ \mathrm{ns}$. The backgrounds from the deuterated matrix and container were measured separately and subtracted by using the proper transmission factors~\cite{richter1992entanglement}. Compared to typical neutron spin-echo experiments where only a few \textit{Q} values are investigated, the present study focuses specifically on the spatial correlations of entangled polymer dynamics. A total of 23 \textit{Q}s are measured in the range of $0.021$--$0.21\ \textup{\AA}\mathrm{^{-1}}$. This allows us to map out the normalized intermediate scattering function $S(Q,t)/S(Q)$ on a $29\times23$ spatiotemporal grid of $29$ discrete correlation times and $23$ $Q$ values.

\section{Spatial correlations from molecular dynamics simulations}\label{sec:MDresults}
\begin{figure}[htp]
    \centering
    \includegraphics[width=\columnwidth]{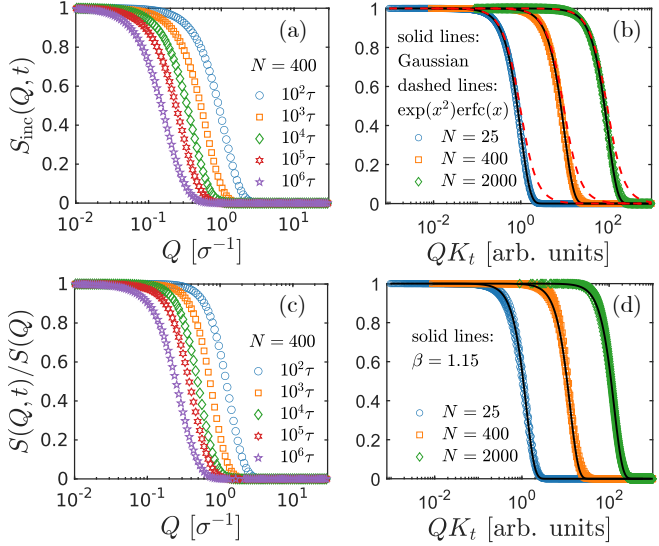}
    \caption{(a) Spatial dependence of $S_{\mathrm{inc}}(Q,t)$ at different correlation times for the $N=400$ semi-flexible system. Panel (b) shows the master curves for the $N=25$, $400$, and $2000$ semi-flexible systems, constructed by horizontally shifting the $S_{\mathrm{inc}}(Q,t)$ data at different correlation times by time-dependent scale factors $K_t$. For clarity, we choose the shift factors in such a way that the master curves of the three systems stay apart from each other. Solid lines in (b): Gaussian function. Dashed lines: $\exp(x^2)\mathrm{erfc}(x)$. Panels (c) and (d) are the results for the normalized single-chain dynamic structure factor $S(Q,t)/S(Q)$, presented in the same format as (a) and (b). Solid lines in (d): compressed Gaussian functions $S(Q,t)/S(Q)=\exp\big[-(Q^2\xi_t^2)^{\beta}/6\big]$ with $\beta=1.15$.}
    \label{fig:MDresults}
\end{figure}

This section presents the results from our molecular dynamics simulations. Fig.~\ref{fig:MDresults}a shows the $S_{\mathrm{inc}}(Q,t)$ of the $N=400$ melt at correlation times both shorter and longer than the entanglement time $\tau_{\mathrm{e}}\approx 2.2\times 10^3\tau$. At first glance, there is no obvious change of spatial dependence of $S_{\mathrm{inc}}(Q,t)$ across $\tau_{\mathrm{e}}$. By applying a horizontal shift factor $K_t$ to each curve, we are able to collapse the data at different times onto a master curve (Fig.~\ref{fig:MDresults}b), which can be approximated by a Gaussian function: $S_{\mathrm{inc}}(Q,t)\approx \exp\big(-Q^2\xi_t^2/6\big)$. The $N=25$ and $2000$ systems behave in a similar way. Additionally, for the two entangled melts at $t>\tau_{\mathrm{e}}$, the spatial decay of $S_{\mathrm{inc}}(Q,t)$ differs substantially from the prediction of Eq.~(\ref{eq:functionForm}). Turning our attention to the normalized single-chain dynamic structure factor $S(Q,t)/S(Q)$, we find similar results (Figs.~\ref{fig:MDresults}c and \ref{fig:MDresults}d), with the exception that the spatial correlations are now described by a ``compressed'' Gaussian function:
\begin{equation}
    \frac{S(Q,t)}{S(Q)}\approx \exp\left[-\frac{1}{6}(Q^2\xi_t^2)^{\beta}\right],\label{eq:compressedGaussian}
\end{equation}
where $\beta \ge 1$ and $\xi_t$ is a characteristic length that is dependent on $t$. For our current simulations, the curves are slightly compressed, with $\beta\approx 1.15$. As we shall show below, these observations are distinctly different from the predictions of several theoretical models.

\begin{figure}[htp]
    \centering
    \includegraphics[width=\columnwidth]{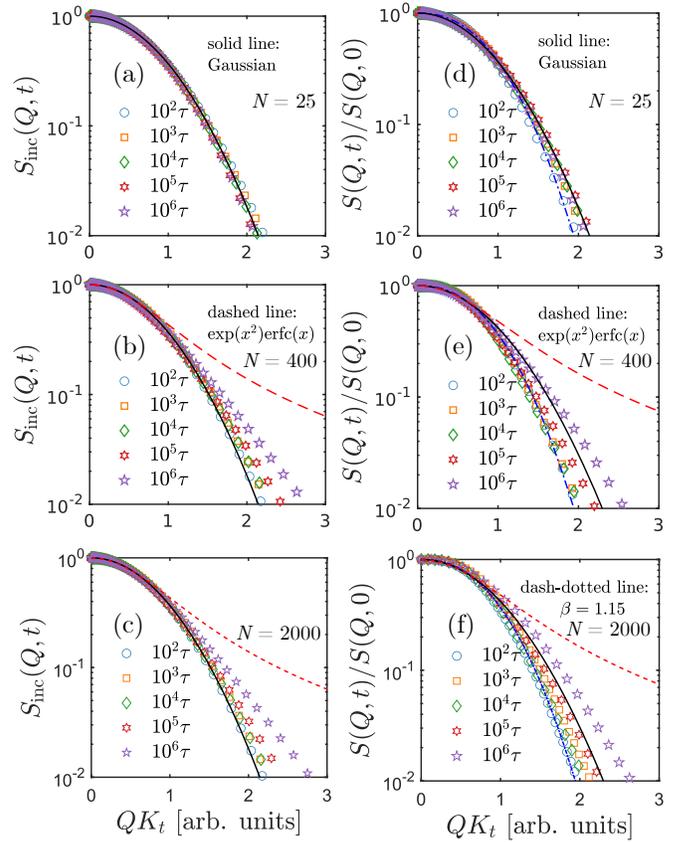}
    \caption{(a)-(c) Tails of the spatial correlations of the incoherent dynamic structure factors $S_{\mathrm{inc}}(Q,t)$ for the $N=25$, $400$, and $2000$ semi-flexible systems. (d)-(f) Results for the normalized single-chain dynamic structure factor $S(Q,t)/S(Q)$. Black solid lines: Gaussian functions. Red dashed lines: $\exp(x^2)\mathrm{erfc}(x)$. Blue dash-dotted lines: compressed Gaussian functions with $\beta = 1.15$.}
    \label{fig:MD_tail}
\end{figure}

Careful inspection shows some subtle changes of spectrum shape as we vary the correlation time. Indeed, the master curves in Fig.~\ref{fig:MDresults} are \textit{approximates} and a closer examination of the spatial correlations reveals important details. Fig.~\ref{fig:MD_tail} presents the same simulation data on a log-linear scale. To highlight the tail of the spatial correlation, we match the initial decays at different times by choosing proper horizontal shift factors $K_t$. The spatial correlation of $S_{\mathrm{inc}}(Q,t)$ for the unentangled melt ($N=25$) remains Gaussian at all times. By contrast, as the correlation time increases above $\tau_{\mathrm{e}}$, the $S_{\mathrm{inc}}(Q,t)$ of the two entangled melts develops a long tail (Figs.~\ref{fig:MD_tail}b and \ref{fig:MD_tail}c). Such behavior clearly arises from topological constraints: localization of real-space density correlations spreads out in the reciprocal space. However, the degree of confinement is evidently much weaker than the prediction of the classical tube model [Eq.~(\ref{eq:functionForm})], where the 3D displacement in the laboratory frame is assumed to be the result of strictly 1D diffusion~\cite{fatkullin1995theory}.

On the other hand, the spatial decay of $S(Q,t)/S(Q)$ follows a slightly compressed Gaussian function ($\beta\approx 1.15$) at short time and a Gaussian function at long time in the unentangled melt (Fig.~\ref{fig:MD_tail}d). For the entangled systems, the spatial correlations eventually become broader than the Gaussian at $t\gg\tau_{\mathrm{e}}$ (Figs.~\ref{fig:MD_tail}e and \ref{fig:MD_tail}f), which resembles the behavior of $S_{\mathrm{inc}}(Q,t)$. Combining the analyses from Figs.~\ref{fig:MDresults} and \ref{fig:MD_tail}, we conclude that the topological constraints have a relatively weak influence on the \textit{functional forms} of spatial distribution of density correlations and the entanglement effect manifests as a long tail in $S_{\mathrm{inc}}(Q,t)$ and $S(Q,t)/S(Q)$.

\begin{figure}[htp]
    \centering
    \includegraphics[width=\columnwidth]{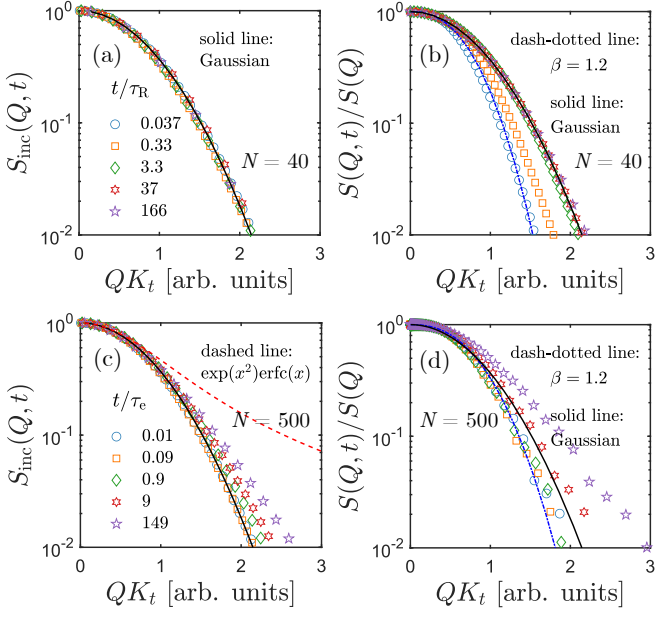}
    \caption{Tails of the spatial correlations for the fully flexible bead-spring model. (a) and (b): results for the unentangled, $N=40$ system. (c) and (d): results for the entangled, $N=500$ system.}
    \label{fig:repulsiveModel}
\end{figure}

All these aforementioned qualitative features of spatial correlations of dynamics are further confirmed by the fully flexible chain simulations. Fig.~\ref{fig:repulsiveModel} shows the spatial correlations of the $N=40$ and $500$ systems. To highlight the tails of the correlations, the data are horizontally shifted and presented on a log-linear scale. Since the entanglement length $N_{\mathrm{e}}$ for this fully flexible bead-spring model falls into the range of 60--86~\cite{Hoy2009topological,Cao2015simulating,grest2016communication}, the $N=40$ chain is unentangled and the $N=500$ system is moderately entangled. Overall, the spatial correlations in these fully flexible chains are qualitatively similar to those in the semi-flexible chains: for the unentangled system, the incoherent scattering function $S_{\mathrm{inc}}(Q,t)$ is always a Gaussian function and the normalized single-chain dynamic structure factor $S(Q,t)/S(Q)$ is a slightly compressed Gaussian function at $t<\tau_{\mathrm{R}}$ and becomes Gaussian when the center-of-mass diffusion takes over at $t\gg\tau_{\mathrm{R}}$; for the entangled system, both the $S_{\mathrm{inc}}(Q,t)$ and $S(Q,t)/S(Q)$ develop a long tail at $t>\tau_{\mathrm{e}}$. Therefore, the main features of the spatial correlations of dynamics reported in this work are indeed a result of topological constraints and insensitive to the details of interchain or intrachain interactions.

\section{Spatial correlations from neutron spin-echo spectroscopy}\label{sec:NSEresults}
\begin{figure}
    \centering
    \includegraphics[width=\columnwidth]{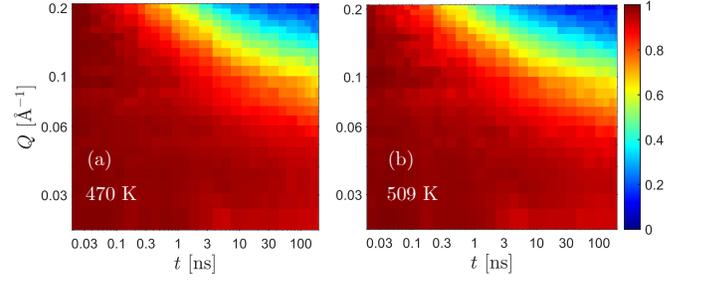}
    \caption{Two-dimensional color maps of the normalized intermediate scattering function (single-chain dynamic structure factor) $S(Q,t)/S(Q)$ obtained from the NSE experiments. (a) 470 K. (b) 509 K. The data were collected on a $29\times23$ spatiotemporal grid of $29$ discrete correlation times and $23$ $Q$ values.}
    \label{fig:NSE_2D_plots}
\end{figure}

\begin{figure}
    \centering
    \includegraphics[width=\columnwidth]{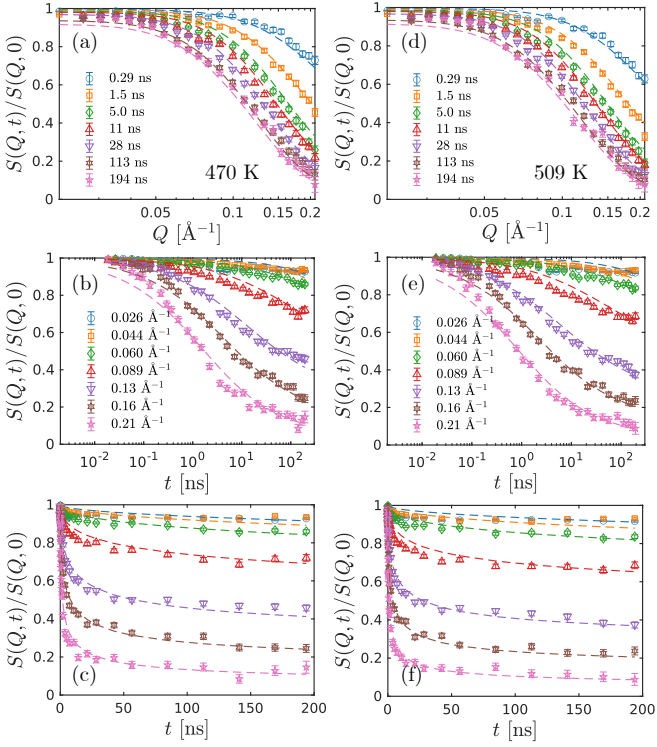}
    \caption{(a) $S(Q,t)/S(Q)$ from the NSE experiments at 470~K, presented at different correlation times. (b) $S(Q,t)/S(Q)$ at different wavenumbers. (c) The same result in (b) presented on linear-linear scale. Panels (d)-(f) are the same plots for the results at 509~K. Dashed lines: fits using Eq.~(\ref{eq:tubeSqt}). The fitting parameters are: $Wb^4=9.92\ \textup{\AA}\mathrm{^4/ps}$, $a=51.8\ \textup{\AA}$, and $\tau_{\mathrm{d}}=14.4\ \mu\mathrm{s}$ for 470~K; $Wb^4=18.7\ \textup{\AA}\mathrm{^4/ps}$, $a=53.1\ \textup{\AA}$, and $\tau_{\mathrm{d}}=13.7\ \mu\mathrm{s}$ for 509~K.}
    \label{fig:NSE_discrete_Q_t}
\end{figure}

\begin{figure}
    \centering
    \includegraphics[width=\columnwidth]{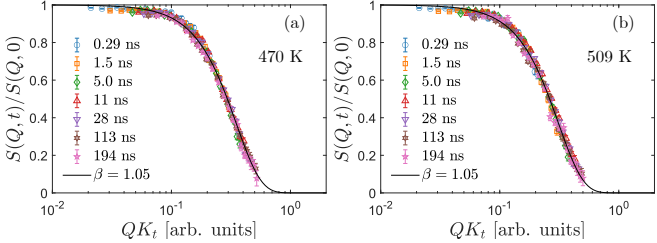}
    \caption{Master curves for NSE $S(Q,t)/S(Q)$ data collected at different correlation times. (a) Results at 470~K. (b) 509~K. Solid lines: compressed Gaussian functions with $\beta=1.05$.}
    \label{fig:NSE_mastercurve}
\end{figure}

\begin{figure}[htp]
    \centering
    \includegraphics[width=\columnwidth]{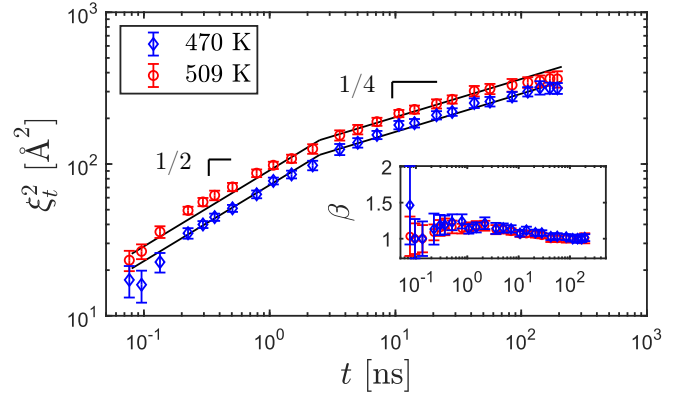}
    \caption{Characteristic displacement $\xi^2_t$ obtained by fitting the spatial decay by Eq.~(\ref{eq:compressedGaussian}). Blue diamonds: results at 470~K. Red circles: results at 509~K. Inset: corresponding shape parameters $\beta$.}
    \label{fig:NSE_results}
\end{figure}

Having examined the spatial correlations of polymer dynamics in molecular dynamics simulations of coarse-grained bead-spring models, we now turn our attention to neutron spin-echo experiments. The normalized intermediate scattering functions $S(Q,t)/S(Q)$ are shown in Fig.~\ref{fig:NSE_2D_plots} in the form of two-dimensional color maps for $T=470$~K and $509$~K. For further quantitative analysis, representative slices of $S(Q,t)/S(Q)$ are presented in Fig.~\ref{fig:NSE_discrete_Q_t} at discrete wavenumbers and correlations times, respectively. In accordance with the standard practice, we perform fits of $S(Q,t)/S(Q)$ by the following tube model equation~\cite{schleger1998clear}:
\begin{eqnarray}\label{eq:tubeSqt}
\frac{S(\Tilde{Q},\Tilde{t})}{S(\Tilde{Q})}&=&\exp\left(-\frac{\Tilde{Q}^2}{6Z}\right)\left[\frac{8}{\pi^2}\sum_{p:\mathrm{odd}}\frac{1}{p^2}\exp\left(-\frac{p^2\Tilde{t}}{3Z}\right)\right]\\*
\nonumber&+&\left[1-\exp\left(-\frac{\Tilde{Q}^2}{6Z}\right)\right]\exp\left(\frac{1}{\pi^2}\Tilde{Q}^4\Tilde{t}\right)\mathrm{erfc}\left(\frac{1}{\pi}\Tilde{Q}^2\Tilde{t}^{1/2}\right),
\end{eqnarray}
where $Z=N/N_{\mathrm{e}}$, $\Tilde{Q}=QR_{\mathrm{g}}$, and $\Tilde{t}=t/\tau_{\mathrm{R}}$, with $R_{\mathrm{g}}$ and $\tau_{\mathrm{R}}$ being the radius of gyration and Rouse relaxation time, respectively. To fully utilize the experimental data, the entire 2D surface in Fig.~\ref{fig:NSE_2D_plots} is fitted by the nonlinear least-squares method. Eq.~(\ref{eq:tubeSqt}) provides a ``fair'' but imperfect description of the experimental data (please see further discussions in Sec~\ref{sec:TheoreticalResults}). In agreement with our simulations, a master curve (Fig.~\ref{fig:NSE_mastercurve}), which is a slightly compressed Gaussian function with $\beta\approx1.05$, can be constructed by applying horizontal shifts to the data in Figs.~\ref{fig:NSE_discrete_Q_t}b and \ref{fig:NSE_discrete_Q_t}e. Note that the entanglement time $\tau_{\mathrm{e}}$ is on the order of a few nanoseconds for this system (Fig.~\ref{fig:NSE_results} and Ref.~\cite{schleger1998clear}). The construction of a master curve for data across $\tau_{\mathrm{e}}$ therefore confirms the weak influence of entanglements on the functional form of spatial correlations.   

Using the compressed Gaussian function [Eq.~(\ref{eq:compressedGaussian})], the characteristic displacement $\xi_t^2$ at each correlation time can be extracted from $S(Q,t)/S(Q)$ for the coherent dynamics (Fig.~\ref{fig:NSE_results}). Curiously, $\xi_t^2$ exhibits a scaling relation of $\xi_t^2\sim t^{1/2}$ at $t\lessapprox 2$~ns and $\xi_t^2\sim t^{1/4}$ at $t\gtrapprox 2$~ns, reminiscent of the well-known behavior of monomer mean-square displacement for the incoherent dynamics of entangled polymers~\cite{de1971reptation,DEbook,kremer1990dynamics,wischnewski2003direct,hsu2016static}. This observed resemblance between collective and self motions is not a coincidence. The similarity between the overall features of $S_{\mathrm{inc}}(Q,t)$ and $S(Q,t)/S(Q)$ (Figs.~\ref{fig:MDresults}, \ref{fig:MD_tail}, and \ref{fig:repulsiveModel}) suggests that the self and pair correlations can be loosely related by a convolution approximation~\cite{vineyard1958scattering} without the de Gennes-Sk\"{o}ld correction~\cite{de1959liquid,skold1967small,skold1967atomic}:
\begin{equation}
    G_{\mathrm{d}}(\mathbf{r},t)\approx \int H(\mathbf{r}-\mathbf{r}^{\prime},t)G_{\mathrm{d}}(\mathbf{r})\,d\mathbf{r}^{\prime},\label{eq:convolution}
\end{equation}
where $H(\mathbf{r},t)=(2\pi)^{-3}\int [S(\mathbf{Q},t)/S(\mathbf{Q})]e^{i\mathbf{Q}\cdot\mathbf{r}}\,d\mathbf{Q}\approx G_{\mathrm{s}}(\mathbf{r},t)$, and $G_{\mathrm{s}}(\mathbf{r},t)$ and $G_{\mathrm{d}}(\mathbf{r},t)$ are the self and distinct parts of the van Hove function, respectively. Here, we use the fact that for high molecular weight polymers, the single-chain structure factor is dominated by pair correlation: $G(\mathbf{r})\approx G_{\mathrm{d}}(\mathbf{r})$. The analysis of $\xi_t^2$ also demonstrates that while the functional form of spatial correlation is not highly sensitive to entanglements, the spatial decay rate $\xi^2_t$ is. In fact, it is easy to recognize that the monomer mean-square displacement $g_1(t)$ is simply the second moment of $G_{\mathrm{s}}(\mathbf{r},t)$. When $G_{\mathrm{s}}(\mathbf{r},t)$ can be approximated by a Gaussian function (Fig.~\ref{fig:MDresults}b): $G_{\mathrm{s}}(\mathbf{r},t)\approx[3/(2\xi_t^2\pi)]^{3/2}\exp[-3\mathbf{r}^2/(2\xi_t^2)]$, we have $g_1(t)=\int\mathbf{r}^2G_{\mathrm{s}}(\mathbf{r},t)\,d\mathbf{r}=\xi_t^2$ and $S_{\mathrm{inc}}(\mathrm{Q},t)=\exp(-\mathbf{Q}^2\xi_t^2/6)$. On the other hand, our preceding analysis indicates that the difference between the functional forms of $G_{\mathrm{s}}(\mathbf{r},t)$ in the entangled and unentangled regimes is quite subtle.

\section{Comparison with theoretical models}\label{sec:TheoreticalResults}
Having discussed the results from molecular dynamic simulations and NSE experiments, we now confront the question of whether the spatial correlations of entangled polymer dynamics can be adequately described by the existing theoretical models. In addition to the classical tube theory [Eqs.~(\ref{eq:tubeSinc}) and (\ref{eq:tubeSqt})], we include in our investigation the slip-spring model of Likhtman~\cite{likhtman2005single} as well as the models of Ronca~\cite{ronca1983frequency} and des Cloizeaux~\cite{desCloizeaux1993dynamic}. We organize the discussions into two subsections: incoherent dynamics (Section~\ref{sec:incoherent}) and single-chain coherent dynamics (Section~\ref{sec:coherent}). For the convenience of the reader, we will give a brief overview of our main findings at the beginning of each subsection.

\subsection{Incoherent dynamics}\label{sec:incoherent}
This subsection compares the incoherent intermediate scattering functions $S_{\mathrm{inc}}(Q,t)$ from the molecular dynamics simulations with those given by three theoretical models: the Rouse model~\cite{rouse1953theory}, the standard tube model~\cite{de1971reptation,DE1,deGennes1981coherent,DEbook,fatkullin1995theory}, and the slip-spring model~\cite{likhtman2005single}. The Rouse model predicts, to the first approximation, Gaussian spatial correlations of incoherent dynamics on both short and long time scales. This gives rationales for the behavior of unentangled polymer melts from MD simulations, as well as that of entangled melts at $t<\tau_{\mathrm{e}}$. For the entangled dynamics, \textit{i.e.}, $S_{\mathrm{inc}}(Q,t)$ at $t\ge\tau_{\mathrm{e}}$, the strict 1D diffusion idea of the classical tube model~\cite{fatkullin1995theory} [Eq.~(\ref{eq:tubeSinc})] creates a confinement that is too strong to be consistent with the simulations (Figs.~\ref{fig:MDresults}b, \ref{fig:MD_tail}b, \ref{fig:MD_tail}c, and \ref{fig:repulsiveModel}c). On the other hand, the slip-spring model~\cite{likhtman2005single} relaxes the impenetrable tube constraint by trapping a Rouse chain with linear springs anchored in space. Our analysis shows such a treatment significantly improves the prediction of the tube model by suppressing the spread of correlations in the reciprocal space, and good agreement is found between the molecular dynamics results and the slip-spring simulations with comparable degree of entanglement. The details of the analyses and calculations are provided below.

\subsubsection{Rouse model}
While the focus of the current study is entangled polymers, it is instructive to first examine the spatial correlations of Rouse dynamics~\cite{rouse1953theory}. According to the Rouse model, the incoherent intermediate scattering function $S_{\mathrm{inc}}(Q,t)$ can be approximated by the following formula on time scales shorter than the Rouse relaxation time $\tau_{\mathrm{R}}$~\cite{de1967quasi}:
\begin{equation}
    S_{\mathrm{inc}}(\Tilde{Q},\Tilde{t})\approx\exp\left(-\frac{2}{\pi^{3/2}}\Tilde{Q}^2\Tilde{t}^{1/2}\right),\label{eq:Rouse_Sinc}
\end{equation}
where $\Tilde{Q}=QR_{\mathrm{g}}$ and $\Tilde{t}=t/\tau_{\mathrm{R}}$. Eq.~(\ref{eq:Rouse_Sinc}) implies that the spatial correlation of $S_{\mathrm{inc}}(Q,t)$ at a given correlation time $t$ can be described by a Gaussian function, $S_{\mathrm{inc}}(Q,t)=\exp\left(-\frac{1}{6}Q^2\xi_t^2\right)$, where $\xi_t$ is a characteristic length scale that depends on $t$. At $t\gg\tau_{\mathrm{R}}$, the incoherent intermediate scattering function is dominated by center-of-mass diffusion, and its spatial correlation is also Gaussian: $S_{\mathrm{inc}}(Q,t)\approx \exp(-Q^2Dt)$, with $D$ being the center-of-mass diffusion coefficient. The Rouse model thus offers an explanation for the Gaussian-like spatial correlations of unentangled polymer dynamics in the MD simulations, and that of entangled polymers at $t<\tau_{\mathrm{e}}$.

\subsubsection{Tube model}
\begin{figure}[h]
    \centering
    \includegraphics[width=0.9\columnwidth]{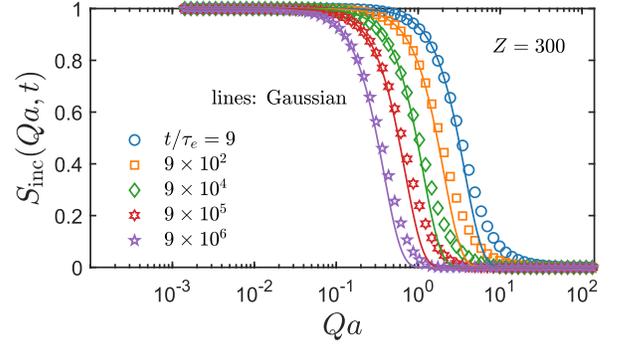}
    \caption{Example of the incoherent scattering function $S_{\mathrm{inc}}(Q,t)$ predicted by the standard tube model~\cite{fatkullin1995theory}. Symbols: tube model calculations for a well-entangled system with $Z=300$. Lines: Gaussian functions.}
    \label{fig:tube_Sinc}
\end{figure}

In Section~\ref{sec:MDresults}, we show that the standard tube model fails to capture the incoherent intermediate scattering function from the MD simulations. A more quantitative analysis can be performed as follows. For $t\ge \tau_{\mathrm{e}}$, the incoherent scattering function $S_{\mathrm{inc}}(Q,t)$ of an entangled polymer can be calculated according to the tube model as~\cite{fatkullin1995theory}:
\begin{equation}
    S_{\mathrm{inc}}(Q,t)= \exp[Q^4y(t)]\mathrm{erfc}[Q^2\sqrt{y(t)}]\exp(-Q^2Dt),\label{eq:tubeSinc2}
\end{equation}
where $D$ is the center-of-mass diffusion coefficient, $\mathrm{erfc}(x)$ is the complementary error function, and $y(t)=a^2\langle s^2(t)\rangle/72$, with $a$ being the tube diameter and $\langle s^2(t)\rangle$ the mean-square curvilinear segment displacement. For a given correlation time $\tau_{\mathrm{e}}\le t \ll \tau_{\mathrm{d}}$, $\exp(-Q^2Dt)\approx 1$ and the spatial decay of $S_{\mathrm{inc}}(Q,t)$ is of the form $\exp(x^2)\mathrm{erfc}(x)$, where $x = Q^2C_t^2$, with $C_t$ a constant determined by $t$. Using the standard result for $\langle s^2(t)\rangle$~\cite{fatkullin1995theory}, we give an example of calculation of $S_{\mathrm{inc}}(Q,t)$ in Fig.~\ref{fig:tube_Sinc} for a well-entangled system with $Z=300$. As implied by Eq.~(\ref{eq:tubeSinc2}), the incoherent scattering function $S_{\mathrm{inc}}(Q,t)$ deviates from the Gaussian function and exhibits a long tail due to the ``hard'' confinement of the tube. This predicted deviation, however, is too strong in comparison with the behavior in molecular dynamics simulations (Figs.~\ref{fig:MDresults}b, \ref{fig:MD_tail}b, \ref{fig:MD_tail}c, and \ref{fig:repulsiveModel}c).

\begin{figure}[h]
    \centering
    \includegraphics[width=\columnwidth]{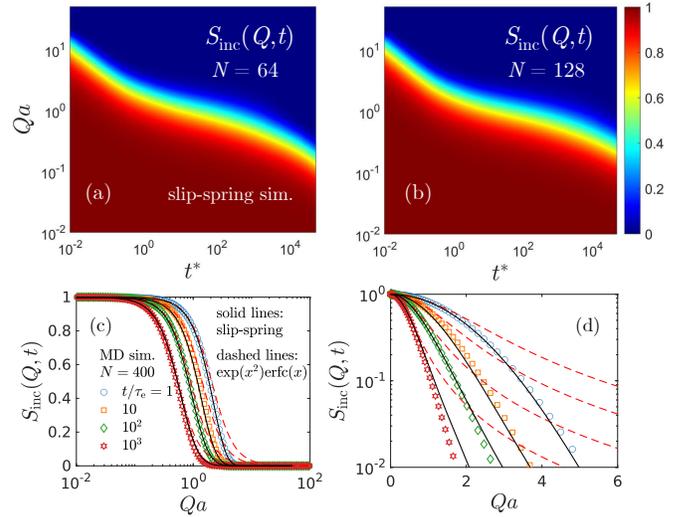}
    \caption{(a) and (b): Spatiotemporal maps of incoherent scattering function $S_{\mathrm{inc}}(Q,t)$ for slip-spring simulations of two different chain lengths $N=64$ and $128$ with $N_{\mathrm{s}}=1/2$, $N_{\mathrm{e}}=4$, and $\zeta_{\mathrm{s}}/\zeta=0.1$. (c) Comparison of molecular dynamics simulations and slip-spring Brownian dynamics simulations. The MD result shown here is for the semi-flexible chain system with $N=400$ and $N/N_{\mathrm{e}}\approx 14$. The BD result is for the $N=64$ system with $N/N_{\mathrm{e}}=16$. The spatial correlations are compared at four normalized correlation times: $t/\tau_{\mathrm{e}}=1$, $10$, $10^2$, and $10^3$. To match the MD and BD results, a tube diameter of $4.1\sigma$ is used to normalize the wavenumber of the MD simulation. Symbols: MD simulations. Solid lines: slip-spring simulations. Dashed lines: $\exp(x^2)\mathrm{erfc}(x)$. (d) The sample result presented on log-linear scale.}
    \label{fig:slipspring_2D}
\end{figure}

\subsubsection{Slip-spring simulations}
To further evaluate the idea of the tube model, we proceed to analyze the incoherent intermediate scattering function of the slip-spring model~\cite{likhtman2005single}, where entanglements take the form of slip-links. Examples of 2D color maps of $S_{\mathrm{inc}}(Q,t)$ are shown in Figs.~\ref{fig:slipspring_2D}a and \ref{fig:slipspring_2D}b. Figs.~\ref{fig:slipspring_2D}c and \ref{fig:slipspring_2D}d compare the spatial correlations of $S_{\mathrm{inc}}(Q,t)$ from the molecular dynamics and slip-spring simulations. The semi-flexible chain model with $N=400$ has an average degree of entanglement of $N/N_{\mathrm{e}}\approx 14$. On the other hand, the $N/N_{\mathrm{e}}$ for the $N=64$ slip-spring system is $16$. After normalizing the MD data with proper entanglement time $\tau_{\mathrm{e}}$ and tube diameter $a$, a good agreement is found between the MD and BD results for $S_{\mathrm{inc}}(Q,t)$. Therefore, by relaxing the topological constraint from an ``impenetrable'' tube to slip-links, the slip-spring model significantly improves the prediction for the incoherent scattering function. This is reflected in the tail of the spatial correlations. Compared to a ``rigid'' tube, the softer slip-spring constraint causes less spread of spatial correlations in the reciprocal space. We note that the slip-spring model includes the constraint release effect through virtual coupling of slip-links, which should also contribute to the improvement of the theoretical prediction.



\subsection{Coherent dynamics}\label{sec:coherent}
In this subsection, we examine the spatial correlations of single-chain coherent dynamics predicted by several theoretical models: the Rouse model, Ronca model, des Cloizeaux model, standard tube model, and slip-spring model. Compared to the incoherent dynamics, the situation for coherent dynamics is much more complicated. The main findings are as follows.

Our analysis of the Rouse model~\cite{rouse1953theory} reveals that it predicts a highly compressed Gaussian function with $\beta\approx1.5$ for $S(Q,t)/S(Q)$ at $t\ll\tau_{\mathrm{R}}$. By contrast, our simulations and NSE experiments indicate that the spatial correlations of high-frequency local dynamics follow a slightly compressed Gaussian function, with $\beta$ on the order of 1.1. Our observations are consistent with the previous report on the failure of the Rouse model~\cite{paul1998chain}, where a weaker-than-expected $Q$ dependence was also found. At $t\le\tau_{\mathrm{e}}$, the slip-spring model, Ronca model, and des Cloizeaux model all produce the normal Rouse dynamics with spatial correlations of a strongly compressed Gaussian function ($\beta\approx1.4-1.5$), at odds with simulations and experiments. Since Rouse dynamics is an essential building block of these models, its failure significantly complicates the discussion of entangled dynamics at $t\ge \tau_{\mathrm{e}}$. The Ronca model incorporates the entanglement effect by adding to the equation of motion a memory term that is related to the relaxation modulus of the system. As a result, the predicted spatial correlations of dynamics is \textit{exactly} Rouse-like even in the entangled regime. Both the slip-spring and des Cloizeaux models envision a long tail for the spatial correlations of $S(Q,t)/S(Q)$ at $t\ge\tau_{\mathrm{e}}$. Overall, the confinement effects introduced in the two models appear too strong, as the functional form of the spatial correlations changes drastically from a highly compressed Gaussian to a ``stretched'' Gaussian function as the correlation time is varied across $\tau_{\mathrm{e}}$. Lastly, our dynamic spatial correlation analysis offers an explanation to the apparent success of Eq.~(\ref{eq:tubeSqt})~\cite{schleger1998clear}. For a well-entangled polymer at $t\gg\tau_{\mathrm{e}}$, the contribution of the local reptation, \textit{i.e.}, the second term on the RHS of Eq.~(\ref{eq:tubeSqt}), is negligibly small. In this limit, the spatial correlations of dynamics are described by a Gaussian function, which is in close agreement with simulations and NSE experiments. This feature distinguishes Eq.~(\ref{eq:tubeSqt}) from the Ronca and des Cloizeaux models. Nevertheless, our preceding analyses show that Gaussian-like spatial correlations are in fact not a genuine trait of reptation. Additionally, Eq.~(\ref{eq:tubeSqt}) suffers a number of intrinsic problems~\cite{likhtman2005single}.

The details of our analyses are presented below.

\begin{figure}[htp]
    \centering
    \includegraphics[width=\columnwidth]{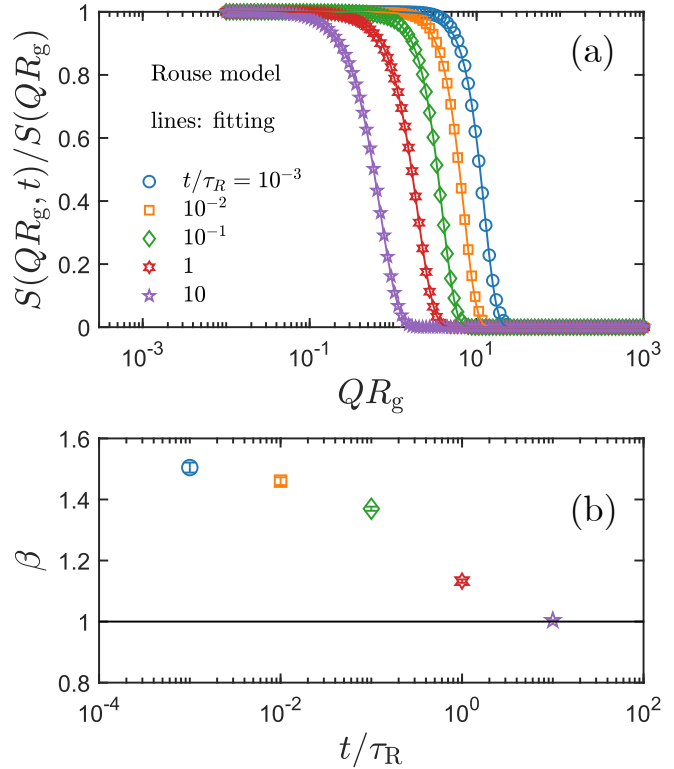}
    \caption{Spatial correlations of coherent single-chain Rouse dynamics. (a) Decay of the normalized single-chain dynamic structure factor at different correlation times. Solid lines: fitting results by the compressed Gaussian function, $S(Q,t)/S(Q)=\exp\big[-(Q^2\xi^2_t)^{\beta}/6\big]$, with $\beta \ge 1$. (b) Dependence of the exponent $\beta$ on the correlation time $t/\tau_{\mathrm{R}}$.}
    \label{fig:Rouse}
\end{figure}

\subsubsection{Rouse model}
The coherent single-chain dynamic structure factor of a Rouse chain of length $N$ is given by~\cite{de1967quasi,DEbook}:
\begin{eqnarray}\label{eq:Rouse_Sqt}
    S(\Tilde{Q},\Tilde{t})&=&\frac{1}{N^2}\exp\bigg(-\frac{2}{\pi^2}\Tilde{Q}^2\Tilde{t}\bigg)\sum_{m,n=1}^{N}\exp\bigg\{\\*\nonumber
    &-&\frac{\Tilde{Q}^2}{N}|n-m|-\frac{4\Tilde{Q}^2}{\pi^2}\sum_{p=1}^{N-1}\frac{1}{p^2}\cos\bigg(\frac{p\pi n}{N}\bigg)\cos\bigg(\frac{p\pi m}{N}\bigg)\\*\nonumber
    &&\big[1-\exp\big(-p^2\Tilde{t}\big)\big]\bigg\},
\end{eqnarray}
where $\Tilde{Q}=QR_{\mathrm{g}}$ and $\Tilde{t}=t/\tau_{\mathrm{R}}$. And the corresponding static single-chain structure factor is:
\begin{equation}
    S(\Tilde{Q})=\frac{1}{N^2}\sum_{m,n=1}^{N}\exp\left(-\frac{\Tilde{Q}^2}{N}|n-m|\right).\label{eq:Rouse_Sq}
\end{equation}
In the limit of $N\to \infty$, Eq.~(\ref{eq:Rouse_Sq}) becomes the well-known Debye function~\cite{DEbook}. The normalized single-chain dynamic structure factor $S(\Tilde{Q},\Tilde{t})/S(\Tilde{Q})$ can be computed by combining Eqs.~(\ref{eq:Rouse_Sqt}) and (\ref{eq:Rouse_Sq}). In the long-time limit, \textit{i.e.}, $t\gg\tau_{\mathrm{R}}$, it is easy to show that $S(\Tilde{Q},t)/S(\Tilde{Q})\approx \exp\big(-2\Tilde{Q}^2\Tilde{t}/\pi^2\big)$. In general, a direct analysis of the spatial correlations through analytical methods is challenging. However, the study can be done numerically. Fig.~\ref{fig:Rouse} presents the spatial correlations of $S(\Tilde{Q},\Tilde{t})/S(\Tilde{Q})$ at different correlation times. For $t\ll \tau_{\mathrm{R}}$, the spatial decay of $S(\Tilde{Q},\Tilde{t})/S(\Tilde{Q})$ is a Gaussian function with $\beta\approx1.5$. As the correlation time increases, $S(\Tilde{Q},\Tilde{t})/S(\Tilde{Q})$ is less compressed and eventually becomes Gaussian at $t\gg\tau_{\mathrm{R}}$. The strong spatial dependence predicted by the Rouse model at $t\ll\tau_{\mathrm{R}}$ is inconsistent with the existing results from molecular dynamics simulations (Figs.~\ref{fig:MDresults}, \ref{fig:MD_tail}, and \ref{fig:repulsiveModel}) and neutron spin-echo spectroscopy experiments (Fig.~\ref{fig:NSE_mastercurve}). This problem complicates the discussion of entanglement effect in many theoretical models, where Rouse-like dynamics serves an essential building block. Lastly, we note that the validity of our numerical evaluations of Eq.~(\ref{eq:Rouse_Sqt}) has been confirmed by Brownian dynamics simulations of the Rouse model.

\subsubsection{Ronca model}
\begin{figure}
    \centering
    \includegraphics[width=\columnwidth]{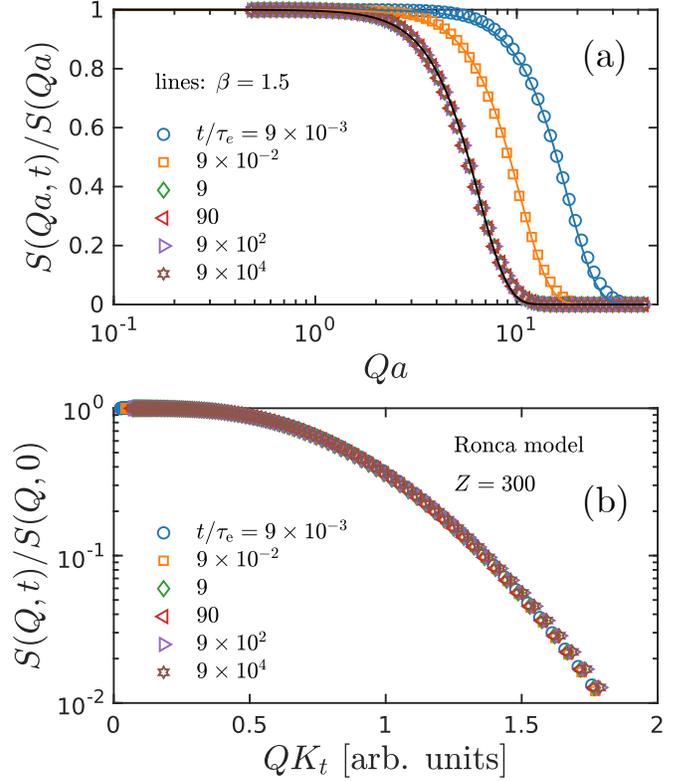}
    \caption{(a) Prediction of the Ronca model~\cite{ronca1983frequency} for the normalized single-chain dynamic structure factor $S(Q,t)/S(Q,0)$ of a well-entangled polymer with $Z=300$. The wavenumber $Q$ is normalized by the tube diameter $a$. Lines: compressed Gaussian functions with $\beta=1.5$. (b) Master curve of $S(Q,t)/S(Q)$ at different correlation times.}
    \label{fig:RoncaModel}
\end{figure}

Historically, a viable candidate for describing the coherent single-chain dynamics of entangled polymers is the model proposed by Ronca~\cite{ronca1983frequency}. The Ronca model enjoyed some success in describing the early neutron spin-echo experiments~\cite{ewen1997neutron}, but ultimately was deemed inferior to the tube model when long Fourier time NSE data became available~\cite{schleger1998clear}. The weakness of the Ronca model can be better appreciated from the perspective of spatial correlations of dynamics. According to the model, the normalized single-chain dynamic structure factor $S(Q,t)/S(Q)$ of an entangled polymer can be calculated as:
\begin{equation}
    \frac{S(\Tilde{Q},\Tilde{t})}{S(\Tilde{Q})}=\frac{\Tilde{Q}^2}{4Z}\int_0^{\infty}\exp\left[-\frac{\Tilde{Q}^2}{8Z}g\left(s,\frac{16}{\pi^2}Z^2\Tilde{t}\right)\right]\,ds,\label{eq:Ronca}
\end{equation}
where $\Tilde{Q}^2=Q^2R_{\mathrm{g}}^2=Q^2a^2Z/6$, $\Tilde{t}=t/\tau_R$, and
\begin{eqnarray}
\nonumber g(x,y) &=& 2x - \exp(x)\mathrm{erfc}\bigg(\sqrt{y}+\frac{x}{2\sqrt{y}}\bigg)\\*
\nonumber &+&\exp(-x)\mathrm{erfc}\bigg(\frac{x}{2\sqrt{y}}-\sqrt{y}\bigg).
\end{eqnarray}
In the limit of $t\to\infty$, the spatial correlation in the entanglement plateau region is
\begin{equation}
    \frac{S(\Tilde{Q},\infty)}{S(\Tilde{Q})}=\frac{\Tilde{Q}^2}{4Z}\int_0^{\infty}\exp\left\{-\frac{\Tilde{Q}^2}{4Z}[s+\exp(-s)]\right\}\,ds.
\end{equation}
The approximation given by Ronca is: $S(\Tilde{Q},\infty)/S(\Tilde{Q})\approx 1-\frac{9}{124Z^2}\Tilde{Q}^4$, which has a much stronger $Q$ dependence than the Gaussian function. Numerically, we find that the $S(Q,t)/S(Q)$ at different correlation times can be approximated by a strongly compressed Gaussian function with $\beta=1.5$. Fig.~\ref{fig:RoncaModel} shows that according to the Ronca model the functional form of $S(Q,t)/S(Q)$ is exactly the same for $t<\tau_{\mathrm{e}}$ and $t>\tau_{\mathrm{e}}$. This behavior is a direct result of the basic approach of the model: the entanglement effect is introduced to the equation of motion by including a memory term that is closely related to the relaxation modulus of the system. Since there is no concept of \textit{topological} constraint, the predicted spatial correlations are still Rouse-like in the entanglement plateau region, with $\beta=1.5$. By contrast, both the molecular dynamics simulations (Figs.~\ref{fig:MDresults}, \ref{fig:MD_tail}, and \ref{fig:repulsiveModel}) and neutron spin-echo spectroscopy experiments (Fig.~\ref{fig:NSE_mastercurve}) indicate that the spatial correlations of entangled dynamics can be approximated by a slightly compressed Gaussian function, with $\beta$ on the order of 1.1. 

\subsubsection{des Cloizeaux model}
Another entangled polymer model of interest is the one proposed by des Cloizeaux~\cite{desCloizeaux1993dynamic}, in which the normalized single-chain dynamic structure factor is described by the following equation:
\begin{eqnarray}\label{eq:desCloizeaux}
    \frac{S(\Tilde{Q},\Tilde{t})}{S(\Tilde{Q})}&=&\frac{Z}{\Tilde{Q}^2}\ln\left(1+\frac{1}{Z}\Tilde{Q}^2\right)\\*\nonumber
    &+&\int_0^{\infty}\mathrm{d}\mu\,e^{-\mu}F\left(\frac{1}{\sqrt{\pi}}\Tilde{Q}^2\Tilde{t}^{1/2},\mu\frac{1}{Z}\Tilde{Q}^2\right),
\end{eqnarray}
with 
\begin{equation*}
    F(x,y)=\int_{0}^{1}\mathrm{d}A\int_{0}^{1}\mathrm{d}B\,\left\{y\exp\left[-yA-G(A,B,x,y)\right]\right\},
\end{equation*}
\begin{eqnarray}
    \nonumber &G&(A,B,x,y)=\frac{1}{\sqrt{\pi}}\int_{0}^{x}\mathrm{d}w\times\\*\nonumber
    &\sum&_{p=-\infty}^{\infty}\left\lvert\exp\left[-\frac{(A-2p)^2y^2}{w^2}\right]-\exp\left[-\frac{(B-2p)^2y^2}{w^2}\right]\right\rvert.
\end{eqnarray}
In the long-time limit, $S(\Tilde{Q},\infty)/S(\Tilde{Q})$ is of the form:
\begin{equation}\label{eq:desCloizeaux2}
    \frac{S(\Tilde{Q},\infty)}{S(\Tilde{Q})}=\frac{Z}{\Tilde{Q}^2}\ln\left(1+\frac{1}{Z}\Tilde{Q}^2\right)-\frac{\Tilde{Q}^2}{2Z}\int_0^1\frac{du\,\ln\lvert2u-1\rvert}{(1+u\Tilde{Q}^2/Z)^2}.
\end{equation}
It is easy to verify that the spatial decay of $S(Q,t)/S(Q)$ predicted by Eq.~(\ref{eq:desCloizeaux2}) is much slower than a Gaussian function, which is inconsistent with our simulations and experiments.

\begin{figure}
    \centering
    \includegraphics[width=\columnwidth]{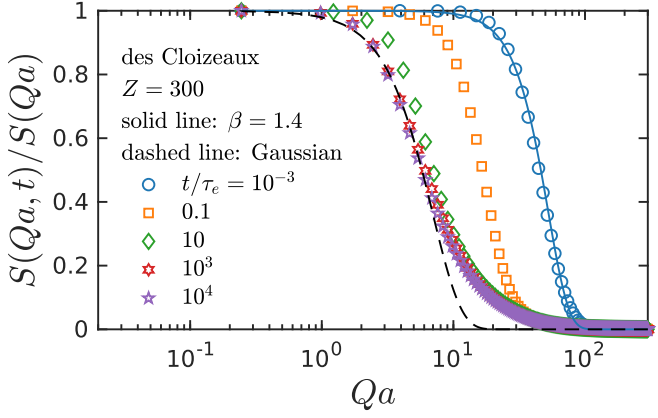}
    \caption{Prediction of the des Cloizeaux model~\cite{desCloizeaux1993dynamic} for the normalized single-chain dynamic structure factor $S(Q,t)/S(Q)$ of a well-entangled polymer with $Z=300$. Solid line: compressed Gaussian function with $\beta=1.4$. Dashed line: Gaussian function.}
    \label{fig:otherModels}
\end{figure}

\subsubsection{Tube model}
In Section~\ref{sec:NSEresults}, we show that the widely used the tube model formula [Eq.~(\ref{eq:tubeSqt})] provides a fair description of the normalized single-chain dynamic structure factor $S(Q,t)/S(Q)$ from neutron spin-echo spectroscopy, in agreement with the previous experiments~\cite{schleger1998clear}. This apparent success of the tube model requires further examination and explanation. For large $Z$ and $t\gg \tau_{\mathrm{e}}$, Eq.~(\ref{eq:tubeSqt}) is dominated by the reptational diffusion term:
\begin{equation}
    \frac{S(\Tilde{Q},\Tilde{t})}{S(\Tilde{Q})}\approx \exp\left(-\frac{\Tilde{Q}^2}{6Z}\right)\left[\frac{8}{\pi^2}\sum_{p:\mathrm{odd}}\frac{1}{p^2}\exp\left(-\frac{p^2\Tilde{t}}{3Z}\right)\right],
\end{equation}
whose spatial dependence is described by a Gaussian function. Furthermore, if $t\ll\tau_{\mathrm{d}}$, $S(\Tilde{Q},\Tilde{t})/S(\Tilde{Q})\approx \exp(-\Tilde{Q}^2/6Z)$. On the other hand, our molecular dynamics simulations and NSE experiments show that the spatial correlations of $S(Q,t)/S(Q)$ can be approximated by a slightly compressed Gaussian function, with $\beta\approx1.1$. Therefore, the effectiveness of Eq.~(\ref{eq:tubeSqt}) comes from the term $\exp(-\Tilde{Q}^2/6Z)$, instead of local reptation dynamics. In the literature, $\exp(-\Tilde{Q}^2/6Z)$ is sometimes referred to as the form factor of the tube~\cite{zorn2002inelastic}. However, a close inspection of the calculations by de Gennes~\cite{deGennes1981coherent} and Likhtman~\cite{likhtman2005single} indicates that this form factor is in fact not an integral part of the classical tube theory: in de Gennes' treatment, the Gaussian tube ``form factor'' is assumed, not derived; in Likhtman's derivation, the tube ``form factor'' is $1/(1+\Tilde{Q}^2/6Z)$, which approaches the Gaussian function only when $\Tilde{Q}^2/6Z\ll 1$.

In addition to Likhtman's criticisms about Eq.~(\ref{eq:tubeSqt}) in Ref.~\cite{likhtman2005single}, here we point out two more problems from the viewpoint of spatial correlation analysis. First, the spatial correlation of the single-chain dynamic structure factor $S(Q,t)/S(Q)$ is not properly normalized for a moderately entangled polymer at $t\gg\tau_{\mathrm{e}}$: $\lim_{Q\to0}S(Q,t)/S(Q)\not\to 1$ in the zero-angle limit. This problem can be seen in the fitting curves of Fig.~\ref{fig:NSE_discrete_Q_t}. Second, as we alluded to in the preceding discussion, the contribution from the local reptation term, \textit{i.e.}, the second term on the RHS of Eq.~(\ref{eq:tubeSqt}), is negligibly small for a well-entangled system at $t\gg\tau_{\mathrm{e}}$. In other words, it is not possible to use Eq.~(\ref{eq:tubeSqt}) to examine the local reptation mechanism in any meaningful way.

\begin{figure}[h]
    \centering
    \includegraphics[width=\columnwidth]{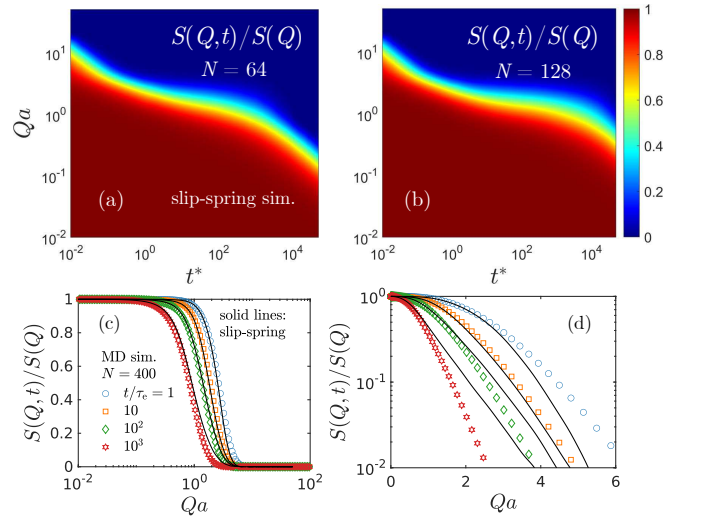}
    \caption{(a) and (b): spatiotemporal maps of normalized single-chain dynamic structure factor $S(Q,t)/S(Q)$ for slip-spring simulations of two different chain lengths $N=64$ and $128$ with $N_{\mathrm{s}}=1/2$, $N_{\mathrm{e}}=4$, and $\zeta_{\mathrm{s}}/\zeta=0.1$. (c) Comparison of molecular dynamics simulations ($N=400$ semi-flexible chains) and slip-spring Brownian dynamics simulations ($N=64$). Symbols: MD simulations. Solid lines: slip-spring simulations. (d) The same result presented on log-linear scale.}
    \label{fig:compare_BD_MD}
\end{figure}

\subsubsection{Slip-spring simulations}
Finally, we consider Likhtman's slip-spring model~\cite{likhtman2005single}, which seems to be able to provide a fair description of incoherent dynamics. Unfortunately, our analysis suggests that the situation is less satisfactory for the single-chain coherent dynamics. An inspection of the spatiotemporal map of single-chain structure factor (Fig.~\ref{fig:compare_BD_MD}) reveals visible changes of color gradient along the wavenumber direction at different correlation times, implying a significant difference in the $Q$-dependence of $S(Q,t)/S(Q)$. This behavior differs qualitatively from the molecular dynamics simulations (Fig.~\ref{fig:MD_illustration}). For $t\le \tau_{\mathrm{e}}$, the slip-spring model predicts a spatial decay of $S(Q,t)/S(Q)$ that is too steep to be consistent with the experiments and simulations (Figs.~\ref{fig:compare_BD_MD}c and \ref{fig:compare_BD_MD}d). This issue comes directly from the failure of the Rouse model. For $t\gg \tau_{\mathrm{e}}$, the long tail predicted by the slip-spring model appears too strong when compared with the molecular dynamics simulations (Fig.~\ref{fig:compare_BD_MD}). 


\section{Relation to previous studies}\label{sec:relations}
\subsection{Molecular dynamics simulations}
\begin{figure}
    \centering
    \includegraphics[width=\columnwidth]{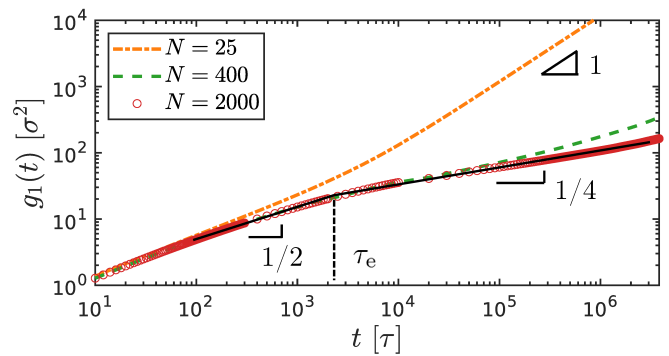}
    \caption{Monomer mean-square displacement $g_1(t)$ for molecular dynamics simulations of semi-flexible chains with different lengths $N=25$, $400$, and $2000$.}
    \label{fig:g1}
\end{figure}

Entangled polymers have been extensively studied by molecular dynamics simulations~\cite{kremer1990dynamics,putz2000entanglement,yamamoto2004entanglements,likhtman2005single,zhou2005primitive,zhou2006direct,likhtman2007linear,bisbee2011finding,likhtman2014microscopic,hsu2016static}. To put this work into perspective, we outline here the key differences between our approach and those adopted by previous investigations. For the self dynamics, the focus of simulation studies is traditionally the mean-square displacement (MSD) of monomers $g_1(t)$ in the laboratory frame, MSD of monomers in the center of mass frame $g_2(t)$, and MSD of the center of mass $g_3(t)$~\cite{kremer1990dynamics}, due to the well-known predictions of the tube model for these quantities~\cite{de1971reptation,DEbook}. As we know, $g_1(t)$ is simply the second moment of the self part of the van Hove function $G_{\mathrm{s}}(\mathbf{r},t)$~\cite{van1954correlations}:
\begin{equation}
    g_1(t)=\int \mathbf{r}^2G_{\mathrm{s}}(\mathbf{r},t)\,d\mathbf{r}.
\end{equation}
The incoherent scattering function $S_{\mathrm{inc}}(\mathbf{Q},t)$, which plays a central role in our current investigation, is nothing but the spatial Fourier transform of $G_{\mathrm{s}}(\mathbf{r},t)$:
\begin{equation}
    S_{\mathrm{inc}}(\mathbf{Q},t)=\int G_{\mathrm{s}}(\mathbf{r},t)e^{-i\mathbf{Q}\cdot\mathbf{r}}\,d\mathbf{r}.
\end{equation}
In other words, this work calls attention to the space-time correlation function $G_{\mathrm{s}}(\mathbf{r},t)$ itself, not just its statistical moments. The monomer mean-square displacement $g_1(t)$ from our simulations is consistent with the previous studies in the literature (Fig.~\ref{fig:g1}): $g_1(t)\sim t^{1/2}$ for $t<\tau_{\mathrm{e}}$ and $g_1(t)\sim t^{1/4}$ for $t>\tau_{\mathrm{e}}$. Interestingly, while the classical tube model correctly predicts the scaling behavior of mean-square displacement, our analysis shows that the functional form of $S_{\mathrm{inc}}(Q,t)$ is not properly described by the theory. This observation underscores the importance of direct analysis of scattering functions, in addition to mean-square displacements. For coherent dynamics, the previous simulation studies follow the traditional strategy and analyze the normalized single-chain dynamic structure factor $S(Q,t)/S(Q)$ at discrete wavenumbers. By contrast, the focus of the current work is to illustrate the benefits of examining the spatial correlations of $S(Q,t)/S(Q)$ at different correlation times.

\subsection{Neutron spin-echo spectroscopy}
The dynamics of entangled polymers has also been investigated extensively by neutron spin-echo spectroscopy over the past several decades~\cite{richter1989microscopic,richter1990direct,butera1991microscopic,richter1992entanglement,richter1993onset,ewen1997neutron,schleger1998clear,wischnewski2002molecular,wischnewski2003direct,zamponi2006molecular}. It is therefore imperative for us to put this work into the context of the existing NSE studies as well. From a technical point of view, our NSE experiment differs from the previous studies in its explicit focus on spatial dependence of dynamics. While a typical study in the past covers only a few \textit{Q}s in a narrow window, the present work examines 23 discrete \textit{Q}s in the range of $0.021$--$0.21\ \textup{\AA}\mathrm{^{-1}}$. Another obvious difference is that instead of analyzing the time correlation of scattering functions at different \textit{Q}s, we place emphasis on the spatial correlations of dynamics at discrete Fourier times.  

\begin{figure}
    \centering
    \includegraphics[width=\columnwidth]{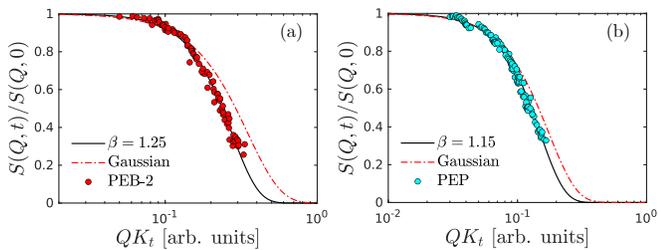}
    \caption{Master curves constructed using our approach for the NSE data reported in (a) Ref.~\cite{schleger1998clear} and (b) Ref.~\cite{wischnewski2003direct}. Symbols: NSE data. Solid lines: compressed Gaussian functions. Dash-dotted lines: Gaussian functions.}
    \label{fig:NSE_literature}
\end{figure}

There is also the question of whether our spatial correlation analysis agrees with the published NSE data on entangled polymers. To properly address this issue, we digitize the polyethylene (PEB-2) data in Ref.~\cite{schleger1998clear} and the polyethylene propylene (PEP) data in Ref.~\cite{wischnewski2003direct}, present the single-chain dynamic structure factor at discrete correlation times, and apply horizontal shift factors to construct master curves. Fig.~\ref{fig:NSE_literature} indicates that both data sets can be described by slightly compressed Gaussian functions, which is in qualitative agreement with our molecular dynamics simulations and NSE experiments. 

In Section \ref{sec:TheoreticalResults} of this paper, dynamic spatial correlation analysis is applied to a number of molecular models of polymers, including the Rouse model~\cite{rouse1953theory}, the classical tube model~\cite{de1971reptation,DEbook,fatkullin1995theory}, the slip-spring model~\cite{likhtman2005single}, the Ronca model~\cite{ronca1983frequency}, and the des Cloizeaux model~\cite{desCloizeaux1993dynamic}. These models were also the subject of previous NSE investigations~\cite{richter1989microscopic,richter1990direct,richter1992entanglement,ewen1997neutron,paul1998chain,schleger1998clear,wischnewski2002molecular,wischnewski2003direct,zamponi2006molecular}, where a major goal was to see if the models provided a good fit to the experimental data. The dynamic spatial correlation analysis, on the other hand, allows us to ask a more critical question: do these models contain all the essential physics? For example, from the viewpoint of spatial correlations, we are able to rationalize the apparent success of the tube model formula [Eq.~(\ref{eq:tubeSqt})] and at the same time understand its deficiencies. Similarly, using dynamic spatial correlation analysis, we are able to clearly illustrate the improvement of the slip-spring model over the original tube model for incoherent dynamics, and identify the missing physics for coherent motions. In this regard, our spatial correlation analysis offers a different perspective for understanding the dynamics of entangled polymers.

\begin{figure}[htb]
    \centering
    \includegraphics[width=\columnwidth]{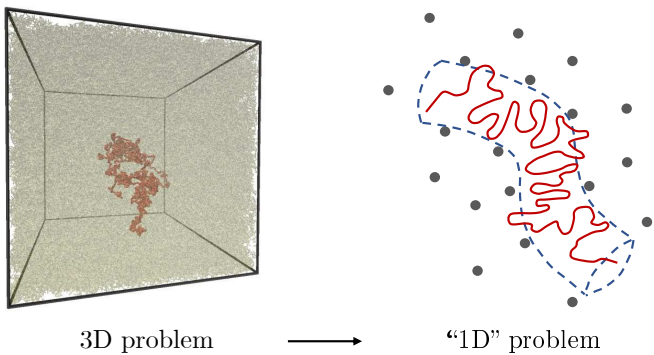}
    \caption{Schematic illustration of the original idea of the tube model: the chain motion is confined in a one-dimensional tube-like region.}
    \label{fig:schematic}
\end{figure}

\section{Concluding Remarks and Summary}\label{sec:summary}
In summary, the dynamics of entangled polymers is examined through the lens of spatial correlation analysis. We find that soft topological constraints have a relatively weak influence on the functional form of spatial correlations. Dynamic localization in real space spreads out in the reciprocal space, appearing as a long tail in the incoherent and coherent intermediate scattering functions. These results cannot be fully described by the theoretical models investigated in this work. To further highlight the lessons we have learned from this exploratory study, we outline below some key implications of the dynamic spatial correlation analysis. 
\begin{itemize}
    \item \textit{A different methodology}: This work calls attention to the spatial correlations of intermediate scattering functions of polymers. For historical and practical reason, intermediate scattering functions of liquids are typically determined and analyzed at a few discrete wavenumbers in scattering experiments or computer simulations. Spatial correlation analysis provides a simple, alternative viewpoint of polymer dynamics. This approach is particularly useful when the traditional time correlation analysis becomes difficult.  
    \item \textit{Universality and signature of entanglement}: Using molecular dynamics simulations and neutron spin-echo spectroscopy, we show that the initial spatial decays of both self and collective dynamics are similar in the unentangled and entangled regimes. This observation holds for different molecular weights and systems of different interactions, implying a level of universality in polymer melt dynamics. The influence of entanglement is mainly reflected in the tail of spatial correlations of intermediate scattering functions, which is difficult to resolve with the current neutron spin-echo technique, but can be readily studied with computer simulations. 
    \item \textit{Theoretical implications}: The Rouse model does not produce proper spatial correlations of collective single-chain dynamics on short time scales ($t<\tau_{\mathrm{R}}$). This is a problem for theoretical modeling of not only unentangled melts, but also entangled polymers, where Rouse-like local motions are elementary steps of chain dynamics. Moreover, the idea of the standard tube model --- strongly confined Rouse dynamics in an impenetrable tube (\textit{i.e.}, strict 1D diffusion idea, see Fig.~\ref{fig:schematic}) --- fails to describe the spatial correlations of incoherent dynamics. The task of describing collective dynamics is even more challenging. The standard tube model formula [Eq.~(\ref{eq:tubeSqt})], the Ronca model [Eq.~(\ref{eq:Ronca})], and the des Cloizeaux model [Eq.~(\ref{eq:desCloizeaux})] all have problems of their own. The slip-spring model, while reasonably successful in describing the segmental self dynamics, still cannot account for the collective spatial correlations of entangled dynamics.  
\end{itemize}

We note that the tube model has long been criticized for its phenomenological treatment of topological constraints, where simplified assumptions about liquid motions are introduced to mimic the entanglement effect. In this regard, dynamic spatial correlation analysis should provide a useful tool for developing a more fundamental theoretical description of polymer entanglement. An intriguing question is whether the essential features of spatial correlations reported here can naturally arise from alternative theoretical approaches to entangled polymers~\cite{schweizer1989microscopic,schweizer1989mode,guenza1999many,yatsenko2004analytical,sussman2011microscopic,sussman2012microscopic,guenza2014localization}. Additionally, our current MD investigations are limited to coarse-grained bead-spring models of polymers. While these simulations are qualitatively consistent with the NSE experiments, quantitative studies with atomistic simulations are highly desirable in the future.

Finally, we offer a speculative insight into possible directions for future theoretical development. Our hypothesis is that incompressibility plays a crucial role in slow dynamics of polymer melts. For a melt consisting of $M$ chains of length $N$, we can define the following normalized dynamic structure factors:
\begin{equation}
    S_{\mathrm{intra}}(\mathbf{Q},t)\equiv\frac{1}{MN^2}\sum_{n}^{M}\sum_{i,j}^{N}\left\langle e^{-i\mathbf{Q}\cdot[\mathbf{R}_{n,i}(t)-\mathbf{R}_{n,j}(0)]}\right\rangle,
\end{equation}
\begin{equation}
    S_{\mathrm{inter}}(\mathbf{Q},t)\equiv\frac{1}{M^2N^2}\sum_{n\neq m}^{M}\sum_{i,j}^{N}\left\langle e^{-i\mathbf{Q}\cdot[\mathbf{R}_{m,i}(t)-\mathbf{R}_{n,j}(0)]}\right\rangle,
\end{equation}
where $S_{\mathrm{intra}}(\mathbf{Q},t)$ and $S_{\mathrm{inter}}(\mathbf{Q},t)$ are respectively the intrachain and interchain dynamic structure factors. $\mathbf{R}_{m,i}(t)$ is the position of segment $i$ in chain $m$ at time $t$. Using the standard argument in the literature, it is straightforward to show that $S_{\mathrm{intra}}(\mathbf{Q},t)$ and $S_{\mathrm{inter}}(\mathbf{Q},t)$ are proportional to each other under the constraint of incompressibility:
\begin{equation}\label{eq:incompressibility}
    S_{\mathrm{intra}}(\mathbf{Q},t) = -MS_{\mathrm{inter}}(\mathbf{Q},t).  
\end{equation}
A derviation of Eq.~(\ref{eq:incompressibility}) is given in the Appendix. Eq.~(\ref{eq:incompressibility}) indicates a close relation between intrachain and interchain dynamics. It is the basis for extracting single-chain dynamic structure factors from NSE experiments on isotopically labeled polymer melts. On the other hand, the static version of Eq.~(\ref{eq:incompressibility}) lays the foundation for determining single-chain structure factors from SANS experiments~\cite{akcasu1980measurement,boue1982convenient,Higgins1994polymers,wang2020quantitative}.

Eq.~(\ref{eq:incompressibility}) implies that the intrachain and interchain dynamics of polymer melts are mirror images of each other due to the incompressibility of liquids. This is a strong constraint for \textit{collective} polymer melt dynamics, which by definition cannot be addressed by single-chain models. Our spatial correlation analysis indicates that the incoherent dynamics of unentangled and entangled polymers can be more or less described by the Rouse and slip-spring models, respectively. However, none of the models examined in this work can give proper predictions for the single-chain coherent dynamics. These observations seem to point to a fundamental difficulty in modeling collective dynamics of polymer melts. In our view, a key missing theoretical ingredient here is an explicit consideration of liquid incompressibility.


\begin{acknowledgments}
The research is supported by the U.S. Department of Energy (DOE), Office of Science, Office of Basic Energy Sciences, Early Career Research Program Award KC0402010, under Contract DE-AC05-00OR22725. The polymer characterization work was performed at Oak Ridge National Laboratory's Center for Nanophase Materials Sciences, which is a DOE Office of Science User Facility. Our computational investigation used resources of the Oak Ridge Leadership Computing Facility at the Oak Ridge National Laboratory, which is supported by the Office of Science of the U.S. Department of Energy under Contract No. DE-AC05-00OR22725. The NSE experiments were performed at the IN15 beamline of the Institut Laue-Langevin (\url{doi.ill.fr/10.5291/ILL-DATA.TEST-3075}). We thank Dr. W.-S. Xu for his help with the molecular dynamics simulations.
\end{acknowledgments}

\appendix*
\section{Role of incompressibility in polymer melt dynamics}
In this appendix, we give a derivation of Eq.~(\ref{eq:incompressibility}) in the main text. Let us consider the coherent scattering function $I(\mathbf{Q},t)$ of a binary blend of hydrogenous and deuterated polymers: $I(\mathbf{Q},t)=b_{\mathrm{H}}^2S_{\mathrm{HH}}(\mathbf{Q},t)+2b_{\mathrm{H}}b_{\mathrm{D}}S_{\mathrm{HD}}(\mathbf{Q},t)+b_{\mathrm{D}}^2S_{\mathrm{DD}}(\mathbf{Q},t)$, where $b_{\mathrm{H}}$ and $b_{\mathrm{D}}$ are the coherent scattering lengths of the hydrogenous and deuterated chain segments respectively, and $S_{\alpha\beta}(\mathbf{Q},t)=\sum_{i=1}^{N_\alpha}\sum_{j=1}^{N_\beta}\langle e^{-i\mathbf{Q}\cdot(\mathbf{r}_i(0)-\mathbf{r}_j(t))} \rangle$ is the partial dynamic structure factor, with $N_\alpha$ and $N_\beta$ being the total numbers of chain segments for species $\alpha$ and $\beta$ respectively. Using the standard technique \cite{roe2000methods}, we can express $S_{\alpha\beta}$ in terms of the fluctuations of the number density $\Delta n_{\alpha}(\mathbf{r},t)\equiv n_{\alpha}(\mathbf{r},t) -\langle n_{\alpha} \rangle$ as:
\begin{equation*}
    S_{\alpha\beta}(\mathbf{Q},t)= \int\int\langle \Delta n_{\alpha}(\mathbf{u},0)\Delta n_{\beta}(\mathbf{r}+\mathbf{u},t)\rangle e^{-i\mathbf{Q}\cdot\mathbf{r}}\,d\mathbf{u}\,d\mathbf{r}, 
\end{equation*}
where the null scattering $\langle n_{\alpha} \rangle \langle n_{\beta} \rangle V (2\pi)^3\delta(\mathbf{Q})$ at $\mathbf{Q}=0$ is discarded. Applying the incompressibility condition $\Delta n_{\mathrm{H}}(\mathbf{r},t)+\Delta n_{\mathrm{D}}(\mathbf{r},t)=0$, we have
\begin{equation*}
    \begin{aligned}
        &S_{\mathrm{HH}}(\mathbf{Q},t)+S_{\mathrm{HD}}(\mathbf{Q},t) =\\ 
        &\int\int\langle\Delta n_{\mathrm{H}}(\mathbf{u},0)[\Delta n_{\mathrm{H}}(\mathbf{r}+\mathbf{u},t)+ \Delta n_{\mathrm{D}}(\mathbf{r}+\mathbf{u},t)]\rangle \\
        &\times e^{-i\mathbf{Q}\cdot\mathbf{r}}\,d\mathbf{u}\,d\mathbf{r}=0.
    \end{aligned}
\end{equation*}
Similarly, we can show that $S_{\mathrm{HD}}(\mathbf{Q},t)+S_{\mathrm{DD}}(\mathbf{Q},t)=0$. It follows that
\begin{equation}\label{eq:A1}
    \begin{aligned}
        I(\mathbf{Q},t) &= (b_{\alpha}-b_{\beta})^2S_{\alpha\alpha}(\mathbf{Q},t) \\
        &=-(b_{\alpha}-b_{\beta})^2S_{\alpha\beta}(\mathbf{Q},t).
    \end{aligned}
\end{equation}
Setting the correlation time $t$ to zero, one recovers the fundamental theorem of small-angle neutron scattering by incompressible liquids \cite{Higgins1994polymers}. For an isotopically labeled polymer melt containing $\phi$ volume fraction of hydrogenous chains and $(1-\phi)$ deuterated chains, the partial dynamic structure factors $S_{\mathrm{HH}}(\mathbf{Q},t)$ and $S_{\mathrm{HD}}(\mathbf{Q},t)$ can be expressed in terms of $S_{\mathrm{intra}}(\mathbf{Q},t)$ and $S_{\mathrm{inter}}(\mathbf{Q},t)$ as:
\begin{equation}\label{eq:A2}
    S_{\mathrm{HH}}(\mathbf{Q},t)=\phi MN^2S_{\mathrm{intra}}(\mathbf{Q},t)+\phi^2M^2N^2S_{\mathrm{inter}}(\mathbf{Q},t),
\end{equation}
\begin{equation}\label{eq:A3}
    S_{\mathrm{HD}}(\mathbf{Q},t)=(1-\phi)\phi M^2N^2S_{\mathrm{inter}}(\mathbf{Q},t).
\end{equation}
Combining Eqs.~(\ref{eq:A1}), (\ref{eq:A2}), and (\ref{eq:A3}), we arrive at a key relation between intrachain and interchain dynamics:
\begin{equation}\label{eq:A4}
    S_{\mathrm{intra}}(\mathbf{Q},t) = -MS_{\mathrm{inter}}(\mathbf{Q},t).  
\end{equation}

\end{document}